\documentclass[11pt,draftcls,onecolumn]{IEEEtranwc}
\usepackage{color}
\usepackage{amsmath,amssymb,amscd}
\usepackage{subfigure, epsfig}

\newtheorem{theorem}{Theorem}
\newtheorem{definition}[theorem]{Definition}
\newtheorem{example}[theorem]{Example}
\newcommand{\mb}{\mathbf}
\newcommand{\ul}{\underline}

\newcommand{\mc}{\mathcal}

\newcommand{\ds}{\displaystyle}

\ifCLASSINFOpdf
 \else
\fi

\begin{document}

\title{Methodologies for Analyzing Equilibria in Wireless Games}

\author{S. Lasaulce, M. Debbah, and E. Altman\\
lasaulce@lss.supelec.fr, merouane.debbah@supelec.fr,
eitan.altman@sophia.inria.fr} \maketitle

%*********************************************************************
\begin{abstract}
Under certain assumptions in terms of information and models,
equilibria correspond to possible stable outcomes in conflicting or
cooperative scenarios where rational entities interact. For wireless engineers, it is of paramount importance to
be able to predict and even ensure such states at which the network
will effectively operate. In this article, we provide non-exhaustive
methodologies for characterizing equilibria in wireless games in
terms of existence, uniqueness, selection, and efficiency.
\end{abstract}

%\begin{IEEEkeywords}
%Game theory, large systems, MAC, MIMO, Nash equilibrium, power
%allocation games, random matrix theory.
%\end{IEEEkeywords}

%*********************************************************************
%*********************************************************************
%*********************************************************************
%\section{Other tentative titles} \label{sec:intro}
%
%\begin{itemize}
%    \item On the methodology of studying equilibria in wireless
%    games
%    \item Keeping his equilibrium in wireless games
%    \item Equilibria in lower layers wireless games
%    \item \textbf{Any other suggestion?}
%\end{itemize}

%*********************************************************************
%*********************************************************************
%*********************************************************************
\section{Introduction}

The major works by Von Neumann, Morgenstern and Nash are recognized
as real catalyzers for the theory of games which originates from the
works by Waldegrave (1713), Cournot (1838), Darwin (1871), Edgeworth
(1881), Zermelo (1913), Borel (1921) and Ville (1938). Whereas the
strong developments of game theory and information theory occured
approximatively at the same time of history, namely in the middle of
the 20th century with the major works by Von Neumann, Morgenstern,
Nash, and Shannon, it is only recently that deepened analyses have
been conducted, at a significant scale, to apply game theory to
communications problems. During the fifty years following the
seminal work by Shannon there were only a small number of papers
adopting a game-theoretic view of communication problems. To cite a
few of them we have \cite{blachman-ire-1957} where point-to-point
communications are seen as a game between the channel encoder
(choosing the best input distribution in terms of mutual information
-MI-) and channel (choosing the worse transition probability in
terms of MI), \cite{berger-it-1971} where source coding is seen as a
game between the source encoder and switcher (modifying the source
distribution) having antagonistic objectives in terms of distorsion,
\cite{trimble-phd-1972} where the author exploits game theory for
the joint signal-and-detector design using game-theoretic techniques
to perform multi-parameter optimization, or also \cite{ericson-1986}
where a legal encoder-decoder pair fights against a jammer. By
contrast, many papers exploiting game theory for communications and
especially wireless communications have been released over the past
fifteen years (\cite{grandhi-allerton-1992}, \cite{altman-ima-1993},
\cite{altman-igtr-2000}, \cite{goodman-pcomm-2000},
\cite{yu-jsac-2002}, etc.) and the phenomenon seems to gain more and
more momentum. There are many reasons for this craze for game theory
in the wireless community, here we will give a few technical reasons
for this. Note that, in this article, we will focus on technical
problems arising at the physical and medium access layers of a wireless network and not
on economic aspects related to it, like the auction problem for
spectrum, even though it is also an important scenario where game
theory is used.

An important reason for this surge of interest in game theory in the area
of wireless communications is the determinant role of certain major
wireless actors like spectrum regulators. Here is an example illustrating
this role: by allowing
anybody to use some portions of the radio spectrum, which are
referred to as unlicensed bands, spectrum regulators have naturally
created a game. Wireless devices (or groups of devices) using these
bands interact with each other by exploiting the available spectral
resources. An other scenario of paramount importance where game
theory is naturally the dominant paradigm to analyze interactions
between devices consists of networks of terminals equipped with a
cognitive radio. Indeed, as spectrum congestion has become a more
and more critical issue, the Federal Communications Commission has
released an important report \cite{report-fcc-2002b} providing a
legal framework for deploying such networks. Another factor for the
great interest of the wireless community in game theory is that
considering a terminal as an intelligent entity capable of observing
and reacting sufficiently rapidly has become a more and more
realistic assumption with the significant progresses of signal
processing (e.g., spectrum sensing algorithms and dramatic increase
of admissible computational complexity). Also, for a long time,
wireless networks were based on the single-user channel paradigm
introduced by Shannon but networks have evolved and multi-user
channel models are now considered as a key element for increasing
reliability, rates and security of wireless communications. In such
networks, terminals share common resources like energy, power,
routes, space, spectrum, time, etc, which implies potential
interaction between them and game theory is, by definition, the
branch of mathematics studying interaction between intelligent
entities. Roughly speaking, the born of multiuser networks appears
to be a milestone delineating two eras: the era were game theory was
used to \emph{analyze} or interpret communication systems and the
era where it is used to \emph{design} or construct communication
system.

When inspecting the works applying game theory to wireless networks,
there is generally a quite significant part of them dedicated to
analyzing the issues related to the notion of equilibrium. This is
not surprising since this concept is instrumental for wireless
engineers. For instance, the existence of an equilibrium can allow them
to predict the effective operating state(s) of the network. Most often, the type of
equilibrium studied is the Nash equilibrium (NE) (see e.g., the
recent papers in \cite{special-issue-jsac-2008}), which corresponds
to the minimum condition of stability for an equilibrium (i.e., it
is robust to a single deviation). Whereas there are many papers
where issues, such as existence and uniqueness of an equilibrium in
a given wireless game, are well treated, there is no paper
attempting to provide a complete methodology
in order to analyze equilibria in wireless
games. It is therefore difficult to have a complete picture of the
existing theorems and techniques available to analyze these
equilibria. One of the purposes of this article is precisely to
contribute to building this picture by giving methodologies for
studying equilibria in non-cooperative wireless games. One of the
main goals of this article is to answer typical questions such as:
Does the network have a pure NE (Sec. \ref{sec:existence})? If yes,
is it unique (Sec. \ref{sec:uniqueness})? If it does not exist, what
can be done? On the other hand, if there are several equilibria, how
to select one of them (Sec. \ref{sec:selection})? If the selected
equilibrium is found to be inefficient in a given sense of
efficiency measure, how can the game be modified to improve its
efficiency (Sec. \ref{sec:efficiency})? This article will be
structured (Sec. \ref{sec:existence}--\ref{sec:efficiency})
according to this list of questions. These sections will be preceded
by a section introducing the notion of equilibrium (Sec.
\ref{sec:concept-eq}) and followed by a section showing to what
extent the provided methodologies can be applied to other types of
equilibrium (Sec. \ref{sec:extensions}). Throughout the article,
more emphasis will be made on the methodology rather than attempting
to enumerate all the main results available. For each main issue
addressed (e.g., existence) some useful definitions, theorems and
techniques will be provided and concisely illustrated by a typical
wireless game at the physical or medium access layer.

\section{The concept of equilibrium}
\label{sec:concept-eq}

In this article, we will mainly focus on a certain type of games
namely static strategic (form) non-cooperative games with complete
information (SNG) and finite number of players. Strategic or normal
form games consist of a triplet $\mc{G} = \left(\mc{K},
\{\mc{S}_i\}_{i \in \mc{K}}, \{ u_i \}_{i \in \mc{K}} \right) $
where $\mc{K} = \{1,...,K\}$ is a finite set of players, $\forall i
\in \mc{K}, \ \mc{S}_i$ is the set of strategies of player $i$ and
$u_i$ his utility (payoff) function; concerning the notations we
will often use $\ul{s} = (s_1,...,s_K)$ to refer to a vector and
$\mc{S} = \prod_{i=1}^{K} \mc{S}_i$ to refer to the Cartesian
product of sets. The games are non-cooperative in the sense that
each player/user wants to selfishly maximize his own utility $u_i$
over his strategy set $\mc{S}_i$. The assumption of complete
information means that every player knows the triplet $\mc{G}$.
Further, they are implicitly assumed to be rational in the sense of
Savage \cite{savage-book-1954} i.e., each player does what is best
for him and rationality of the players is assumed to be common
knowledge. Even though we
restrict our attention to SNGs, the methodology provided here can
be, to a large extent, applied to other types of games, which will
be discussed in Sec. \ref{sec:extensions}. Also, most of our
attention will be dedicated to pure-strategy Nash equilibria (pure
NE), which correspond to the minimum condition of stability that is, the NE
strategy profiles are stable to a single deviation. Indeed, a
pure NE is defined as follows:
\begin{definition}[NE] \emph{The vector $\ul{s}^*$ is a
(pure) NE for $\mc{G}$ if $\forall i \in \mc{K}, \forall s_i' \in
\mc{S}_i, \ u_i(s_i^*, \ul{s}_{-i}^*) \geq u_i(s_i', \ul{s}_{-i}^*)$
where we use the notation $\ul{s}_{-i}$ to refer to the strategy profile
of all players except for player $i$.}
\end{definition}
An NE is therefore stable to a single deviation. In some contexts,
for example, in the context of population (large number of players),
a stronger condition of stability can be required. This is the case
of evolutionary stable equilibria, which are stable to the
deviation of a fraction of a population
\cite{maynardsmith-nature-1973}. In the core of this article only
pure, mixed and correlated equilibria will be mentioned. Indeed, one
has to distinguish between pure, mixed, and correlated strategies.
As already seen, for a player, choosing a pure strategy consists in
picking one element in his set of possible actions. Implementing a
mixed strategy consists in associating a probability with each of
the possible actions and run the series of actions generated by the
corresponding lottery. For example, a transmitter can decide to
transmit at full power or not at all, by following the realizations
of a Bernouilli random variable. A mixed Nash equilibrium is
therefore a vector of probability distributions verifying the Nash
property in Def. 1 where $s_i$ represents a distribution, denoted by
$\ul{q}_i$, and $u_i$ has to be replaced with the expected utility
for player $i$: $\tilde{u}_i(\ul{q}_1,...,\ul{q}_K) = \sum_{\ul{s}
\in \mc{S}} p(\ul{s}) u_i(\ul{s})$ where $p(\ul{s}) = \prod_ {j \in
\mc{K}}  q_{j}(s_j) $ and $q_{j}(s_j)$ is the probability with which
player $j$ chooses the action $s_j \in \mc{S}_j$. In contrast with
mixed equilibria, in correlated equilibria \cite{aumann-jme-1974}
the lotteries used by the players can be correlated (by coordination
signals) that is, in general, we do not have $q(\ul{s}) = \prod_i
q_i(s_i)$.

\section{Existence}
\label{sec:existence}

Mathematically speaking, proving the existence of an equilibrium amounts
to proving the existence of a
solution to a fixed-point problem \cite{nash-aca-1950}; this is why so much effort is
made to derive fixed-point theorems (see e.g.,
\cite{border-book-1985}, \cite{smart-book-1974}) by game theorists.
Therefore, it will not be surprising to see that equilibrium
existence theorems are based on topological properties of the
strategy sets of the players and topological and geometrical
properties of their utility. This explains why the
respective works by Lefschetz \cite{lefschetz-ams-1929}, Hopf
\cite{hopf-mz-1929}, Brouwer \cite{brouwer-ma-1912} and Kakutani
\cite{kakutani-dmj-1941} on fixed-point theorems (\textbf{FPT}) have
had an important impact in game theory. To prove the existence of
an equilibrium at a problem at hand, one may
always attempt to derive an FPT, which is the most general method.
However, there are
many scenarios assuming usual channel models and performance metrics, and
for which existing theorems are sufficient. For example, Shannon
transmission rates or rate regions have desirable convexity
properties that are needed in standard known theorems for
equilibrium existence. The
purpose of this section is to provide certain of these useful
existence theorems and mention some examples where they have been
applied in the literature of wireless communications.

A very useful existence theorem is a theorem stated in \cite{fundenberg-book-1991},
resulting from the contributions of \cite{debreu-aca-1952}, \cite{fan-aca-1952}, and
\cite{glicksberg-ams-1952}. The corresponding theorem, which will call
the Debreu-Fan-Glicksberg theorem as in \cite{fundenberg-book-1991}, can be applied
if the users' utilities are quasi-concave and some other properties
are verified. Before stating the corresponding theorem, let us
review the definition of a quasi-concave function: a function $\psi$
is quasi-concave on a convex set $\mc{S}$ if, for all $\alpha \in
\mathbb{R}$, the upper contour set $\mc{U}_{\alpha} = \{x \in
\mc{S}, f(x) \geq \alpha \}$ is convex. We will use the acronym
\textbf{QG} to refer to games in which utilities are quasi-concave.

\begin{theorem}[Debreu-Fan-Glicksberg 1952] \emph{Let $\mc{G}$ be an
SNG. If $\forall i \in \mc{K}$:  $\mc{S}_i$ is a compact and convex
set; $u_i(\ul{s})$ is a continuous function in the profile of
strategies $\ul{s}$ and quasi-concave in $s_i$; then the game
$\mc{G}$ has at least one pure NE.} \label{theo-debreu}
\end{theorem}

A special case of this theorem is when the utility functions are
concave. In this respect, Theorem 1 by Rosen \cite{rosen-eco-1965}
for \emph{concave} $K-$person games can be seen as a corollary of
Theorem \ref{theo-debreu}. Furthermore, if the utility functions are
assumed to be affine functions, Nash existence theorem
\cite{nash-aca-1950} of a mixed equilibrium in \emph{finite} games
(\textbf{FG}; $|\mc{K}| < + \infty$, $|\mc{S}_i| < + \infty$) can
also be seen as a special case of Theorem \ref{theo-debreu}. This is
due to the fact that the mixed strategy of player $i\in \mc{K}$,
i.e., the distribution $q_i$ used by player $i$, belongs to a
compact and convex set $[0,1]$ and his (averaged) utility is an
affine function of $q_i$.
\begin{example} \emph{For the distributed energy-efficient power control
(PC) problem introduced in \cite{goodman-pcomm-2000}, $\mc{G}$ is
defined as follows. The strategy of user $i\in \mc{K}$ is his
instantaneous transmit power $p_i \in [0, P_i^{\mathrm{max}}]$ and
his utility is $u_i(\ul{p}) = \frac{f(\mathrm{SINR}_i)}{p_i}$ where
$f:\mathbb{R}^+ \rightarrow [0,1] $ is a sigmoidal efficiency
function (e.g., the packet success rate) and $\mathrm{SINR}_i$ is
the signal-to-interference plus noise ratio for user $i$. This game
can be shown to be quasi-concave and thus has at least one pure NE.}
\label{ex:noncoop-pc}
\end{example}
Of course, there is no reason why utilities used in wireless games
should always be quasi-concave. For instance, in the case of the
energy-efficient PC game we have mentioned, the quasi-concavity
property is lost when the utility is modified into
$\tilde{u}_i(\ul{p}) = u_i(\ul{p}) + \alpha p_i$, which corresponds
to implementing a linear pricing technique \cite{saraydar-com-2002}.
If a game can be shown to be non quasi-concave, it still can have
some nice properties that ensure the existence of a pure NE. This is
precisely the case if one has to deal with an S-modular game
(\textbf{SMG}) \cite{milgrom-eco-1990}, \cite{topkis-book-1998} or a
potential game (\textbf{PG}) \cite{monderer-geb-1996}. S-modular
games include submodular games \cite{topkis-jco-1979} and
supermodular games \cite{milgrom-eco-1990}, \cite{topkis-book-1998}.
In what follows we give the definition, an existence theorem and an
example of game for these types of games.

\begin{definition}[S-modular games] \emph{The strategic game  $\mc{G}$ is
said to be supermodular (resp. submodular) if: $\forall
i \in \mc{K}$, $\mc{S}_i$ is a compact subset of $\mathbb{R}$; $u_i$
is upper semi-continuous in $\ul{s}$; $\forall i \in \mc{K}, \forall
\ul{s}_{-i} \geq \ul{s}_{-i}'$ the quantity $u_i(\ul{s}) - u_i(s_i,
\ul{s}_{-i}')$ is non-decreasing (resp. non-increasing) in $s_i$.}
\end{definition}

Furthermore if the utilities of the game are twice differentiable
there is a simple characterization of SMGs.

\begin{definition}[Characterization of S-modular games] \emph{If $\forall i \in
\mc{K},$ $u_i$ is twice differentiable, then the game $\mc{G}$  is
submodular (resp. supermodular) if and only if}
\begin{equation}
\forall (i,j) \in \mc{K}^2, i\neq j, \frac{\partial^2 u_i}{\partial
s_i
\partial s_j} \leq 0 \ (\mathrm{resp.} \ \geq 0).
\end{equation}
\end{definition}
An interpretation of a super-modular (resp. sub-modular) game is
that if the other players $-i$ increase (resp. decrease) their
strategy, player $i$ has interest in increasing (resp. decreasing)
his. One of the nice properties of SMGs is that they do not require
convexity, concavity assumptions on the utilities to ensure the
existence of an NE. The following theorem for SMGs can be found in
\cite{topkis-jco-1979} and follows from Tarski fixed-point theorem
\cite{tarski-pjm-1955}.
\begin{theorem}[Topkis 1979] \emph{If $\mc{G}$ is an
SMG, then it has at least one pure NE.}
\end{theorem}

An example of supermodular games can be found in
\cite{saraydar-com-2002} where the authors uses a linear pricing
technique to improve the energy-efficiency of the NE of the
distributed PC game. The corresponding utilities are $\forall i \in
\mc{K}, \tilde{u}_i(\ul{p}) = u_i(\ul{p}) + \alpha p_i$ which are
not quasi-concave, as already mentioned. An example of submodular
games can be found in \cite{yao-qs-1995} and \cite{altman-tac-2003}.

Now we turn our attention to an other important type of games:
potential games. This type of games has been introduced by Monderer
and Shapley in \cite{monderer-geb-1996}.

\begin{definition}[Potential games] \emph{The game $\mc{G}$ is an exact (resp. ordinal)
potential game if there exists a function $\phi$ such that $\forall
i \in \mc{K}, \forall \ul{s} = (s_i, \ul{s}_{-i}) \in \mc{S},
\forall s_i' \in \mc{S}_i$, $u_i(\ul{s}) - u_i(s_i', \ul{s}_{-i}) =
\phi(\ul{s}) - \phi(s_i', \ul{s}_{-i})$ (resp. $u_i(\ul{s}) -
u_i(s_i', \ul{s}_{-i}) > 0 \ \Leftrightarrow \ \phi(\ul{s}) -
\phi(s_i', \ul{s}_{-i}) > 0$).}
\end{definition}
It is important to note that the potential function is independent
of the user index: for every player, $\phi$ allows one to quantify
the impact of a unilateral deviation on all the users' utilities in
exact PGs while it gives the sign of the difference of utilities in
ordinal PGs. Like SMGs we have the following existence theorem.
\begin{theorem}[Monderer-Shapley-1 1996] \emph{If $\mc{G}$ is a PG with a
finite number of players, compact
strategy sets, and continuous utilities, then it
has at least one pure NE.} \label{theo-monderer-shapley}
\end{theorem}

\begin{example} \emph{In \cite{scutari-icassp-2006} the authors formulate a constrained PC problem
as a game where each user wants to: minimize his transmit power $p_i
\in [0, P_i^{\mathrm{max}}]$ subject to the constraint
$f_i(\mathrm{SINR}_i) \geq \gamma_i$. Considering, without loss of
generality $u_i(\ul{p}) = \log(p_i)$ for the users's cost functions,
it can be checked that $\phi(\ul{p}) = \sum_{i=1}^K \log(p_i)$ is a
potential function for this game.}
\end{example}

We see that Theorem \ref{theo-monderer-shapley} is very useful if a
potential function can be found. If such a function cannot be easily
found, other results can be used to check if the game is potential.
As mentioned in \cite{monderer-geb-1996}, a very simple case where
it is to easy to verify if the game is potential is the case where
the strategy sets are intervals of $\mathbb{R}$. In this case we
have the following theorem.

\begin{theorem}[Monderer-Shapley-2 1996] \emph{Let $\mc{G}$ be a game in which
the strategy sets are intervals of real
numbers. Assume the utilities are twice continuously differentiable.
Then $\mc{G}$ is a potential game if and only if}
\begin{equation}
\forall (i,j) \in \mc{K}^2, \frac{\partial^2 (u_i - u_j)}{\partial
s_i
\partial s_j} = 0.
\end{equation}
\end{theorem}

\begin{figure}
  \begin{center}
  \includegraphics[angle=0,width=12cm]{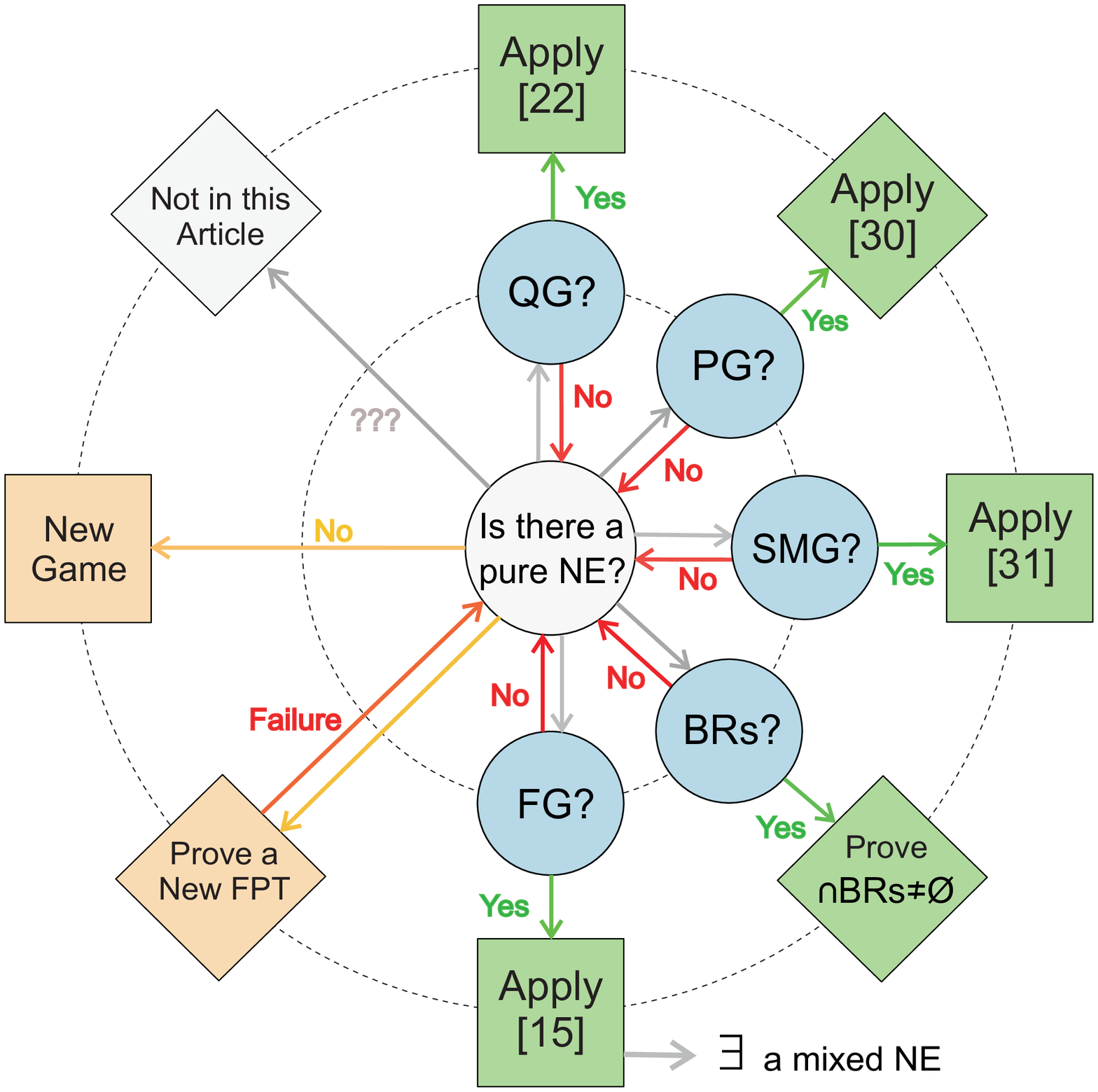}
    \caption{A (non-exhaustive) methodology for proving the existence of a pure Nash equilibrium
    in strategic games. The reader has to refer to the acronyms and
    references used in Sec. \ref{sec:existence}.}
    \label{fig:existence}
\end{center}
\end{figure}

It is interesting to notice that in the (non-exhaustive) list of
theorems stated so far, none of them requires to explicit the best
responses of the players: the best response \textbf{(BR)} of player
$i \in \mc{K}$ corresponds, by definition, to the set of strategies
$\mathrm{BR}_i(\ul{s}_{-i})$ maximizing the utility of user $i$ when
the rest of the world plays $\ul{s}_{-i}$:
$\ds{\mathrm{BR}_i(\ul{s}_{-i}) = \arg \max_{s_i} u_i(s_i,
\ul{s}_{-i})}$. In general $\mathrm{BR}_i$ can be a correspondence
but it is a function in many wireless games addressed in the current
literature of communications. If the BRs can be explicated, the
existence proof boils down to proving that the BRs have a non-empty
intersection, which can be very simple in some scenarios. For
instance in the power allocation (PA) game of
\cite{belmega-gamenets-2009} where the authors study interference
relay channels, the best response of the users are piecewise affine
functions in the case of their amplify-and-forward protocol. The
existence issue and even uniqueness issue are quite simple to
analyze in such a case. To conclude this section, we summarize the
methodology presented in Fig. \ref{fig:existence}. In the box ``Not in this
article'' one could for instance, put the
recent work on distributed power control \cite{brihaye-valuetools-2009}
where the authors show that instead of proving
an FPT (mathematically), the powerful model checking
concept can be used to verify (with
a computer but rigorously) the existence
of a certain network properties or a winning strategy.

%\emph{Conclusion.} In this section we have seen some ways of proving
%the existence of a pure Nash equilibrium. Essentially the
%methodology is as follows: 1. Analyze the properties of the game and
%apply and exploit an existing theorem; 2. Otherwise, come back to
%the definition of an NE by expressing the best responses and
%studying their intersection points; 3. At last, derive a fixed-point
%theorem specific to the game under investigation. Of course, it can
%turn out that the answer is that there is no pure Nash equilibrium.
%In this case, a possible solution is to consider other types of
%equilibrium, for example mixed equilibria or evolutionary equilibria
%[CITE]. The latter type of equilibrium corresponding to large number
%of users, an approximation might have to be made. Another solution
%is to modify the game, for example by changing the utility functions
%[CITE].
%
%Nash, Debreu, Rosen. Link between the
%        theorems: pure and miwed strategies, convexification of what
%        (mixing, correlated equilibrium).\\

\section{Uniqueness}
\label{sec:uniqueness}

Once one is ensured that an equilibrium exists, a natural question
is to know whether it is unique. This is important not only for
predicting the state of the network but also crucial for convergence
issues. Unfortunately, there are not so many general results on the
equilibrium uniqueness. In this section, for sake of clarity, we
will distinguish between two types of situations, depending whether
the BR of every player can be explicated or not.

\subsection{The BRs do not need to be explicated}
\label{sec:BR-no}

A natural question would be to ask whether the Debreu-Fan-Glicksberg
theorem has a counterpart for uniqueness, that is, there exists a
general uniqueness theorem for quasi-concave $K-$player games. To
the best of the author's knowledge, the answer is no. However, there
is a powerful tool for proving the uniqueness of a pure NE when the
players' utilities are concave: this tool is the uniqueness theorem
derived by Rosen \cite{rosen-eco-1965}. This theorem states that if
a certain condition, called diagonally strict concavity (DSC), is
met, then uniqueness is guaranteed. This theorem is as follows.

\begin{theorem}[Rosen 1965] \emph{Assume that $\forall i \in \mc{K}$:  $\mc{S}_i$ is a compact and
convex set; $u_i(\ul{s})$ is a continuous function in $\ul{s} \in
 \mc{S}$ and concave in $s_i$. Let $\ul{r} =
(r_1,...,r_K)$ be an arbitrary vector of fixed positive parameters.
Define the pseudogradient of the function $ w_{\ul{r}} = \ul{r}
\times \ul{u}^T$ by $\ul{\gamma}_{w_{\ul{r}}}(\ul{s}) = \left[r_1
\frac{\partial u_1}{\partial s_1}(\ul{s}), ...,  r_K \frac{\partial
u_K}{\partial s_K}(\ul{s})\right]^T $. If the following condition
(DSC) holds
\begin{equation}
\forall (\ul{s}, \ul{s}') \in \mc{S}^2, \ul{s} \neq \ul{s}': \
(\ul{s}  - \ul{s}') \left(\ul{\gamma}_{w_{\ul{r}}}(\ul{s})  -
\ul{\gamma}_{w_{\ul{r}}}(\ul{s}') \right)^T  > 0
\end{equation}
then the game $\mc{G}$ has a unique NE.}
 \label{theo-2-rosen}
\end{theorem}
To illustrate this theorem we provide an example of wireless game
where it has been applied successfully.

\begin{example} \emph{In \cite{belmega-twc-2009} the authors
generalized the water-filling game of \cite{lai-it-2008} to fast
fading multiple access channels (MACs with one base station -BS- and
$K$ mobile stations -MS-) with multiple antennas both at the
transmitters and receiver (uplink case). In this game, each user
wants to maximize his ergodic transmission rate by choosing the best
precoding strategy. While showing that this game is concave is quite
easy, proving its uniqueness is less trivial. It turns out that the
DSC of Rosen has a matrix counterpart here and can be proved to be
true. Proving the DSC boils down to proving that:
$\mathrm{Tr}\left( \mb{M} \mb{N} + \mb{P} \mb{Q} \right) > 0$
where $\mb{M} = \mb{A} - \mb{B}$, $\mb{N} = \mb{B}^{-1} -
\mb{A}^{-1}$, $\mb{P} = \mb{C} - \mb{D}$, $\mb{Q} =
(\mb{B}+\mb{D})^{-1} -(\mb{A}+\mb{C})^{-1}$, and $\mb{A}$, $\mb{B}$,
 $\mb{C}$, $\mb{D}$ are positive matrices.}
\end{example}

\subsection{When the BRs can be explicated}
\label{sec:BR-yes}

If the BR of every player can be expressed, it is possible to
analyze their properties and for some classes of functions (or
correspondences) to characterize the number of intersection points
between them. A nice class of BRs is the class of standard BRs. Standard
functions have been introduced by Yates in \cite{yates-jsac-1995}; they are defined as
follows.

\begin{definition}[Standard functions] \emph{A vector function
$g: \mathbb{R}_{+}^K \rightarrow \mathbb{R}_{+}^K$ is said to be
standard if it has the two following properties:
\begin{itemize}
\item Monotonicity: $\forall (\ul{x}, \ul{x}') \in
\mathbb{R}_{+}^{2K}$, $\ul{x} \leq \ul{x}' \Rightarrow g(\ul{x})
\leq g(\ul{x}')$;
\item Scalability: $\forall \alpha > 1$, $\forall \ul{x} \in
\mathbb{R}_{+}^K$, $g(\alpha \ul{x}) < \alpha g(\ul{x})$.
\end{itemize}}
\label{def:standard-functions}
\end{definition}

In \cite{yates-jsac-1995} it is shown that if a function is standard
then it has a unique fixed-point. Applying this result to the best responses
of a game we have the following result.

\begin{theorem} [Yates 1995]\emph{Assume that the best responses of the
strategic-form game $\mc{G}$ are standard. Then, the game has a
unique NE.}
\end{theorem}

A simple wireless game where the BRs have these properties is the
energy-efficient PC game introduced by \cite{goodman-pcomm-2000},
which we have already mentioned. The BRs can be shown to be (see
e.g., \cite{saraydar-com-2002}, \cite{lasaulce-twc-2009}): $\forall \in \{1,...,K\}, \
\mathrm{BR}_i(\ul{p}) = \frac{\beta^*}{|h_i|^2} \left(\sigma^2 +
\sum_{j=0}^{i-1} p_j |h_j|^2 \right) $, where $\beta^*$ is a
constant,  $h_i$ the instantaneous channel gain of user $i$ and
$\sigma^2$ the reception noise variance. It can be checked that the
BRs are monotonic and scalable.

As mentioned in the end of Sec. \ref{sec:existence}, if the BRs are
available and quite simple to exploit, it might be possible to find
their \emph{intersection} points. The number of intersection points
corresponds to the number of equilibria. A well-known game where
this kind of approaches is very simple is the Cournot duopoly
\cite{cournot-memoire-1838} for which the BRs of the two players are
affine and intersect in a single point. A counterpart of the Cournot
duopoly in wireless networks is treated, for instance, in
\cite{belmega-icassp-2009}.

%----------------------------------------
\section{Equilibrium selection}
\label{sec:selection}

In fact, there are important scenarios where the NE is not unique.
This typically happens in routing games \cite{milchtaich-an-1996},
\cite{roughgarden-geb-2004} and coordination games
\cite{cooper-book-1998}. Another important scenario, where such a
problem arises, corresponds to games where the choice of actions
from different players is not independent, for instance
non-cooperative games with correlated constraints and generalized NE
\cite{altman-adg-2000} is one of the solution concepts. A central
feature in these constrained games is that they often possess a
large number of equilibria. Natural questions that arise concern the
selection of an appropriate equilibrium. What can be done when one
has to deal with a game having multiple equilibria? Are they some
dominant equilibria? Are there some equilibria fairer and more
stable than others? Obviously, the selection rule is strongly
related to the used fairness criteria. As equilibrium selection is a
theory in itself \cite{harsanyi-book-2003}, the authors will give
here a very incomplete view of the general problem and only partial
answers to these questions. In fact, we will mention only three
issues related to the equilibrium selection problem: how to select
an equilibrium in concave games; the role of the game dynamics in
the selection; the role of efficiency in the selection. These issues
have been chosen here because they are strongly connected to the
content of the other sections of this article.

In the case of concave games, here again, Rosen
\cite{rosen-eco-1965} gives a very neat way to tackle the problem.
In \cite{rosen-eco-1965}, the author introduced the notion of
\emph{normalized equilibria}, which gives a way of selecting an
equilibrium.
\begin{definition}[Normalized equilibrium] \emph{Let $\mc{G}$ be a concave SNG,
$\ul{r}$ a vector of positive parameters, $\mc{C} = \left\{\ul{s} \in \mc{S}, \
h(\ul{s}) \geq 0\right\}$ is a constraint set. An
equilibrium $\ul{s}^*$ of this game is said to be a normalized NE associated with
$\ul{r}$ if there exists a constant $\lambda$ such that
$\lambda_i = \frac{\lambda}{r_i}$ where $\lambda_i$ are the multipliers corresponding
to the Kuhn-Tucker conditions $\lambda_i h(\ul{s}) =0$.}
\end{definition}
This concept has been recently applied by \cite{altman-infocom-2009}
to study decentralized MACs with constraints. The impact of these
constraints is to correlate the players' actions. As mentioned in
the beginning of this section, there can be multiple NE in this type
of games. This is precisely what happens in decentralized MACs. One
of the problems that arises in such contexts is to know how a player
values the fact that constraints of another player are satisfied or
violated. Some extreme cases are: (i) A player is indifferent to
satisfaction of constraints of other players; (ii) Common
constraints: if a constraint is violated for one player then it is
violated for all players. The concept of normalized equilibrium,
applicable to concave games, is precisely a possible way of
predicting the outcome of such a game and/or selecting one of the
possible equilibria. Specifically, the authors of
\cite{altman-infocom-2009} have shown that, in their context of
multiple access game with multiuser detection, the normalized
equilibrium achieves maxmin fairness and is also proportionally fair
(for these notions see e.g., \cite{mo-spie-1998}).

\begin{figure}
  \begin{center}
  \includegraphics[width=12cm]{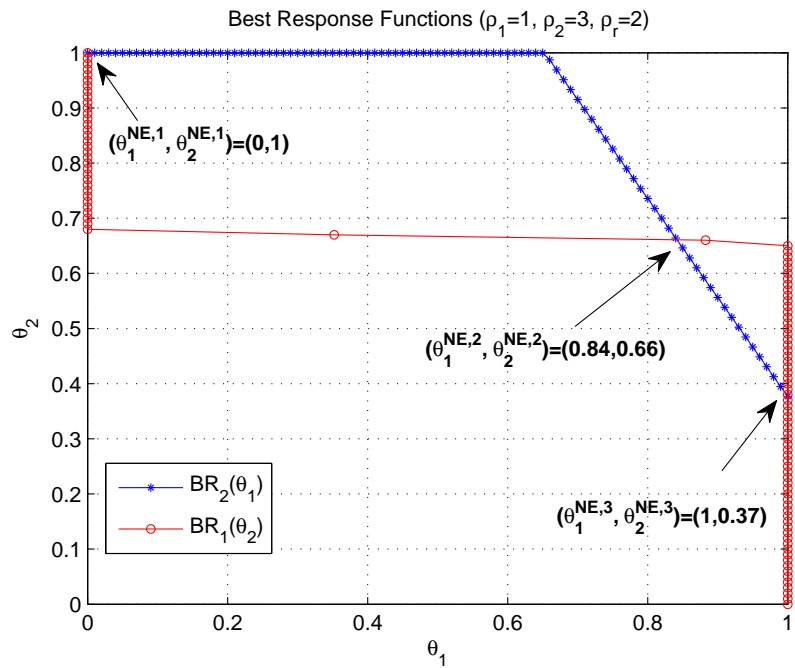}
    \caption{In two-band two-user interference
    relay channels, PA games have multiple NE \cite{belmega-gamenets-2009}. If the users plays and
    observe each other alternately, the game converges to an NE depending on the
      starting point of the game.}
    \label{fig:multiple-NE}
\end{center}
\end{figure}

So far, we have always assumed static games with complete
information namely the game is played in one shot, based on the fact
that every player knows everything about the game. This is in this
precise framework that existence and uniqueness of an NE has been
discussed. Interestingly, the Nash equilibria predicted in such
framework can be observed in others that are less restrictive in
terms of information assumptions. These other frameworks include the
situation where each player observes the actions played by the
others, react to them by playing his BR, the others update their
strategy accordingly, and so on. It turns out that these games can
converge to an NE that would be obtained if the players were knowing
the game completely and playing in one shot.  Fig.
\ref{fig:multiple-NE} shows the possible NE in the PA game of
\cite{belmega-gamenets-2009} in two-band two-user interference relay
channels. In this figure $\theta_i$ represents the power fraction
user $i$ allocates to a frequency band, $1-\theta_1$ being the
fraction allocated to the other band. It can be checked that the
sequence $\{\theta_i^{(0)}; \theta_{-i}^{(1)} =\mathrm{BR}_{-i}(
\theta_i^{(0)});  \theta_i^{(2)} =\mathrm{BR}_i( \theta_{-i}^{(1)});
\theta_{-i}^{(3)} =\mathrm{BR}_{-i}( \theta_i^{(2)}),... \}$ will
converge to one of the three possible NE, depending on the
\emph{game starting point} i.e., on the value of $i\in\{1,2\}$ and
the value of $\theta_i^{(0)} \in[0,1]$. As a more general
conclusion, we see that the initial operating state of a network can
determine the equilibrium state in decentralized networks having
certain convergence properties. We cannot expand on this issue here
but it is important to know that games with standard BRs, PGs, and
SMGs have attractive convergence properties. For example, the authors of
\cite{sastry-tsmc-1994} have shown how simple learning procedures,
based on mild information assumptions, converge to the NE predicted
in the associated game with complete information.

More generally, if there is a certain hierarchy in the game, this
hierarchy can be exploited by one or several players to enforce a
given equilibrium. The desired equilibrium can be selected because
of its \emph{efficiency}. Therefore, equilibrium efficiency is also
a way of selecting an equilibrium. The fact that this equilibrium
will effectively occur depends on whether there exists an entity
capable of influencing the game. In the scenario of
\cite{belmega-icassp-2009} where two point-to-point
communications compete with each other (interference channel), the
network owner chooses the best location for the added relay in order
to maximize the network sum-rate at the equilibrium.

\section{Efficiency}
\label{sec:efficiency}

We have seen that a simple scenario where game theory is a natural
paradigm is the case of decentralized wireless networks. A
well-studied problem is the PC problem in the uplink of MACs. One of
the motivations for decentralized PC in such a context is to
decrease the amount of signaling sent by the BS for PC. But there is
a priori no reason why a decentralized network of partially or
totally autonomous and selfish terminals should perform as well as
its centralized counterpart, for which the PC problem can be
optimized jointly at the BS. This poses the problem of efficiency of
the network. More specifically, for a decentralized network having a
unique NE, it is important to characterize the equilibrium
efficiency since it is the state at which the network will
spontaneously operate. As mentioned above, if there are multiple
equilibria, equilibrium efficiency can be used as a discriminant
factor to select one of them. Whatever the equilibrium be unique or
resulting of a selection, two critical issues arise: how to measure
the network equilibrium efficiency (Sec.
\ref{sec:measure-efficiency});  how to improve its efficiency (Sec.
\ref{sec:improve-efficiency}) when the overall network performance
at the equilibrium is found to be not satisfying.

\subsection{Measuring equilibrium efficiency}
\label{sec:measure-efficiency}

A well-known way of characterizing efficiency of an equilibrium is
to know whether it is Pareto-optimal (PO). Pareto-optimality is
defined as follows.

\begin{definition}[Pareto-optimality] \emph{A profile of strategies
$\ul{s}^{\mathrm{PO}}$ is said to be PO if there is a non-empty subset
of players $\mc{K}_{++}$ such that
\begin{equation}
\forall i \in \mc{K}_{++}, \forall  \ul{s} \in \mc{S}, \
u_i(\ul{s}^{\mathrm{PO}}) > u_i(\ul{s}),
\end{equation}
and
\begin{equation}
\forall i \in  \mc{K} / \mc{K}_{++}, \forall  \ul{s} \in \mc{S}, \
u_i(\ul{s}^{\mathrm{PO}}) \geq u_i(\ul{s}).
\end{equation}}
\end{definition}
Said otherwise, $\ul{s}^{\mathrm{PO}}$ is Pareto-optimal if there
exists no other profile of strategies for which one or more players
can improve their utility without reducing the utilities of the
others. A simple example of Pareto-optimal profile of strategies
is as follows.
\begin{example} \emph{Assume a canonical $2-$user MAC \cite{wyner-it-1974}\cite{cover-book-1975}, $Y
= X_1 + X_2 + Z$ with a receiver implementing successive
interference cancellation (SIC). Consider the very simple game
where, knowing the decoding order used by the receiver, the users
want to maximize their Shannon transmission rate $u_i(\ul{p}) =
\log_2\left(1+ \mathrm{SINR}_i \right)$ by choosing the best value
for their transmit power $p_i \in [0, P_i^{\mathrm{max}}]$. It can
be checked that all the profiles of strategies corresponding to the
full cooperation segment of the capacity region frontier are
Pareto-optimal. In fact, along this segment, we even have a zero-sum
game $\forall \ul{p} \in [0, P_1^{\mathrm{max}}] \times [0,
P_2^{\mathrm{max}}], \ u_1(\ul{p}) + u_2(\ul{p}) =
R_{\mathrm{sum}}$, where $R_{\mathrm{sum}} = \log_2 \left(1 +
\frac{P_1^{\mathrm{max}} +  P_2^{\mathrm{max}}}{\sigma^2} \right)$
is the MAC sum-rate.}
\end{example}
In the example we have considered to illustrate Pareto-optimality we
see that PO profiles of strategies are such that the sum of the
utilities is maximized. This observation corresponds to the following general
result (see e.g., \cite{arora-book-2004}): every strategy profile $\ul{s}$ which maximizes the weighted sum
$\sum_{i \in \mc{K}} \alpha_i u_i(\ul{s}) $ is Pareto-optimal,
with $\alpha_i > 0$. This result is useful to
determine PO profiles of a game. In particular, the results holds when
$\alpha_i =1$. The corresponding function corresponds to a very well-known
quantity: the social welfare. The social welfare of a game is defined as follows
\cite{arrow-book-1963}:
\begin{definition}[Social welfare] T\emph{he social welfare of a game is
defined as the sum of the utilities of all players:
\begin{equation}
w = \sum_{i=1}^{K} u_i.
\end{equation}}
\end{definition}

Social welfare, which corresponds to the average utility of the
players (up to a scaling factor), is a well-known absolute measure
of efficiency of a society, especially in economics. Is this
quantity relevant in wireless communications? In theory, and more
specifically in terms of ultimate performance limits of a network
(Shannon theory), social welfare coincides with the network
sum-rate. In contrast with many studies in economics, we have in
communications, thanks to Shannon theory, a fundamental limit for
the social welfare. For example, if we have $K$ terminals, each of
them implementing a selfish PC algorithm to optimize his Shannon
transmission rate, communicating with two BSs connected with each
other, we know that $w$ cannot exceed the transmission rate of the
equivalent virtual $K\times 2$ multiple input multiple output (MIMO)
system. In practice, it can be a good measure if the players undergo
quite similar propagation conditions, in which case, the utilities
after averaging (e.g., over fading gains) can be close. If the users
experience markedly different propagation conditions the use of
social welfare can be sometimes arguable and even leads to very
unfair solutions. We can mention at least three reasons why social
welfare has to be replaced, in some scenarios, with other measures
of global network performance. First, it is an absolute measure and
therefore does not translate how large the gap between the
performance of the decentralized network and its centralized
counterpart is. Second, as mentioned, it can be unfair. Third, while it
has a very nice physical interpretation when the users' utilities
are chosen to be Shannon transmission rates, its meaning is much
less clear in contexts where other utilities are considered (e.g.,
energy-efficiency). Before providing a way of dealing with the first
drawback mentioned, let us illustrate the second and third ones by
the example of energy-efficient PC games.

\begin{example} \emph{In \cite{lasaulce-twc-2009} the authors have studied, in
particular, a non-cooperative energy-efficient PC game (see Ex.
\ref{ex:noncoop-pc} for more details)) when the BS implements SIC.
When the BS can optimize the decoding order associated with SIC, the
authors have shown that after optimization of $w = \sum_{i=1}^{K}
u_i = \sum_{i=1}^{K} \frac{f(\mathrm{SINR}_i)}{p_i}$, the users who
were ``rich'' in terms of link quality are now even ``richer''. This
shows that $w$ can be an unfair measure of energy-efficiency of the
network. On other hand, the other performance metric considered in
\cite{lasaulce-twc-2009}, the equivalent virtual MIMO system, $v
= \ds{\frac{\sum_{i=1}^{K} f(\mathrm{SINR}_i)}{\sum_{i=1}^{K}
p_i}}$, is both fairer in terms of energy-efficiency and has a more
physical interpretation than $w$.}
\end{example}
Let us go back to the first drawback of $w$. To deal with it,
\cite{papadimitriou-stoc-2001} introduced the concept of price of
anarchy (PoA). The PoA was initially used to measure the
inefficiency of equilibria in non-atomic selfish routing games, that
is with a large number of players. In  the context of games with a
finite number of players, it is defined as follows.
\begin{definition}[Price of anarchy] \emph{The PoA of the game $\mc{G}$ is
equal to the ratio of the highest value of the social welfare (joint
optimization) to the worse NE of the game:
\begin{equation}
\ds{\mathrm{PoA} = \frac{\max_{\ul{s} \in \mc{S}}
w(\ul{s})}{\min_{\ul{s}^* \in \mc{S}^{\mathrm{NE}}} w(\ul{s}^*)}}
\end{equation}
where $\mc{S}^{\mathrm{SE}}$ is set of Nash equilibria of the game.}
\end{definition}
The price of stability (PoS) is defined similarly by replacing the
denominator of the PoA with the best NE of the game, that is PoA and
PoS coincide if there exists a unique NE. Interestingly, these
quantities can be bounded in some cases. For instance, in the case
of non-atomic routing games the PoA can be upper bounded by using
variational inequalities. In a simple scenario where the cost
functions of the players are affine functions this upper bound is
$\frac{4}{3}$ (see e.g., \cite{nrtv-book-2007}).

%Define $T_i$\\
%\cite{arrow-book-1963}, PoA \cite{papadimitriou-stoc-2001}, PoS
%(other related criteria) \cite{legrand-gamecomm-2007}\\

\subsection{Improving equilibrium efficiency}
\label{sec:improve-efficiency}

When the performance of the network at the (considered) NE is found
to be insufficient, the network or the game can be modified. There
are many ways of doing this and we will just mention a few of them.
What is important to have in mind is that the corresponding changes
generally require allocating some resources (time-slots, band, ...)
in order for the nodes to exchange some information and implement
the new strategies.

A possible way of improving the equilibrium efficiency is to
transform the non-cooperative game/network into a cooperative
game/network. Note that we distinguish between cooperative networks
and cooperative games. In a cooperative network, say a cooperative
MAC where transmitters exchange cooperation signals
\cite{Willems-it-may-1983}, \cite{SendonarisI-com-nov-2003}, the
transmitters can still be selfish. In a cooperative game, some
players help each other. For instance, in a \emph{team game}, the
players have a common objective and they do not necessarily exchange
some signals for this purpose. A team formulation of the PC game
introduced in \cite{goodman-pcomm-2000} will however require full
CSI at all the transmitters. Indeed, as mentioned in
\cite{lasaulce-twc-2009}, if all the transmitters want to
maximize $w$, every transmitter will need to know $h_1,...,h_K$, in
contrast with the non-cooperative game where only $h_i$ is required
at transmitter $i$. Of course, it is also possible and even required
in a large network to form smaller groups of players: this is the
principle of \emph{coalitional games} (see e.g.,
\cite{vonneumann-book-1944}, \cite{aumann-bam-1960} for coalitional
games with and without transferable utility respectively). If each
member of a group or coalition has enough information, the coalition
can even form a virtual antenna \cite{saad-globecom-2008} and the
gain brought by cooperation has to be shared between the players
of the group. In such games, we see that we can have both locally
cooperative networks and a non-cooperative game between the
coalitions. In cooperative games, the counterpart of
the NE solution concept for non-cooperative games
is the Shapley value \cite{shapley-book-1953}. Note that there are other forms of cooperation,
sometimes more implicit. This is the case of \emph{repeated games}
(see e.g., \cite{aumann-report-1966} for infinitely repeated games).
Repeated games are a special case of dynamic games which consist in
repeating at each step the same static game and the utilities
results from averaging the static game utilities over time. In such
a games, certain agreements between players on a common plan can be
implemented, and a punishment policy can also be implemented to
identify and punish the deviators. A simple wireless game where the
concept of repeated games has been applied is the water-filling game
by \cite{lai-it-2008} to the fast fading MAC with single-antenna
transmitters and a multi-antenna receiver. The authors show that the
capacity region frontier is reached by repeating the non-cooperative
game where selfish users maximize their transmission rate. At last
we will mention an other form of cooperation: \emph{bargaining}.
Very interestingly, Nash has proved that cooperative games
\cite{vonneumann-book-1944} can be studied with the same concepts as
those used for non-cooperative games: a cooperative profile of
strategies can be obtained as a subgame perfect equilibrium (which is an equilibrium
having a certain robustness against changes in the main
equilibrium plan, see \cite{selten-ijgt-1975} for more details)
resulting from a bargaining procedure between the players. An
example of application of this concept is \cite{larsson-jsac-2008}
where two multi-antenna BSs cooperate by implementing a Nash
bargaining solution. The bargaining solution has the advantage of
being simple but for more than two players, this solution is rarely
used because it does take into account the fact that an isolated
group of players cooperating only with each other can be formed.

The ways of improving the efficiency of the network equilibrium we
have mentioned so far, are generally very demanding in terms of CSI
at the transmitters and can possibly require to establish new
physical links between some nodes. Since cooperation can be costly
in terms of additional resources, a more reasonable solution can be
to merely coordinate the players. Here, we do not mean creating
coordination games \cite{cooper-book-1998}, which corresponds to a
certain type of games we will not study here, but adding a certain
degree of \emph{coordination} in a non-cooperative game.
Coordination between users can be stimulated, for instance, by using
existing broadcasting signals like DVB or FM signals (case of public
information), or by introducing dedicated signals sent by a BS
(which can send both private and public information). The fact that
all players of the game have access to certain (public or/and
private) signals generally modify the players' behaviors. This
knowledge can lead to a more efficient equilibrium, which can be a
new NE or even a correlated equilibrium \cite{aumann-jme-1974}. The
authors of \cite{altman-networking-2006} have shown that, in a MAC
where a slotted ALOHA protocol is assumed, a public coordination
signal induces correlated equilibria that are more efficient in
terms of reducing the frequency of collisions between users. More
generally, it can be shown \cite{aumann-jme-1974} that the set of
achievable equilibria is enlarged by using common and private
messages in the context of correlated equilibria. At last we will
just mention two other usual techniques to improve the performance
of a network at the equilibrium: (1) implementing a \emph{pricing}
technique or (2) introducing a certain degree of \emph{hierarchy} in
the game. Technique (1) has been used by \cite{saraydar-com-2002}
for energy-efficient PC games and by \cite{lai-it-2008} for
transmission rate-efficient PC games. Technique (2) has been used by
\cite{lasaulce-twc-2009} for energy-efficient PC games and by
\cite{basar-infocom-2002} for transmission rate-efficient PA games.

In this section we have mentioned several techniques to deal with
equilibrium inefficiencies that are inherent to the non-cooperative
game formulation. What is the best technique to be used? The answer
to this question depends on many factors. Among the dominant factors
we have the feasibility of the technique, predictability of the
effective network state and of course the performance of the
solution. Feasibility includes for example realistic information
assumptions (CSI), complexity constraints at the terminals, problems
of measurability. Predictability can be the impossibility to prove
the uniqueness of the equilibrium. For example, implementing pricing
necessitates to modify the original utility functions and uniqueness
can be lost or not predictable after theses changes (see e.g.,
\cite{saraydar-com-2002}). Performance can be a certain target in
terms of quality of service.

%Full cooperation\\
%Implicit cooperation\\
%Coordination\\
%Hierarchy\\
%Pricing\\
%Non-coop case, coop case (2 sub-cases)\\
%Natural ways in communication: cooperative\\
%Difference between explicit and implicit cooperation\\
%CSI is important\\

%To select an eq. To improve the eq eff.\\
%How to measure the NE eff: Pareto-efficiency, social efficiency
%\cite{arrow-book-1963}, PoA \cite{papadimitriou-stoc-2001}, PoS
%(other related criteria) \cite{legrand-gamecomm-2007}, fairness,
%discussion on the way to measure the global efficiency
%(hayel-gamecomm-2008, touati-gamecomm-2007, etc) (for example it can
%be easy for Shannon rate-type
%        utilities but it requires much more care for utilities e.g. those
%        based on spectral or energy efficiency
%        \cite{goodman-pcomm-2000}\cite{meshkati-jsac-2006}\cite{hayel-gamecomm-2008}
%        \cite{lasaulce-twc-sub-2008}) ...\\
%
%
%
%\begin{itemize}
%    \item Introduce coordination. Private and common message, correlated
%    eq. It will
%         shown the set of
%        achievable equilibria is enlarged by using common and private messages in the
%         context of correlated equilibria).
%
%    \item Pricing.
%
%    \item Introduce hierarchy.
%
%    \item Cooperation. It can be implicit (no additional physical
%    link), or explicit (cooperation links). Implicit: repeated
%    games, agreements,  Explicit: coalitional games, cooperation channels.
%
%    \item\textbf{ Eitan: other suggestions?}
%
%\end{itemize}

\section{Beyond Nash equilibria in strategic games with finite number of players}
\label{sec:extensions}

In this article we have mainly focused on static/one-shot games with
complete information and finite number of (rational) players. The
methodology described in this article can be, to a large extent,
re-used in other types of games having possibly different types of
equilibrium. Taking into account the space constraints we had, the
authors propose to briefly discuss two assumptions here: 1. The
assumption on the number of players: What does the concept of NE
become in games with a large number of players? An important concept
of solution to these games is Wardrop equilibrium
\cite{wardrop-ice-1952}; 2. The static game assumption: What does
the concept of NE become in dynamic games?

\subsection{Nash versus Wardrop in large games}

To establish the link and differences between the Nash and Wardrop
equilibria in a concrete way, our discussion is built on specific
but useful scenarios in wireless communications, the problem of
packets routing in wireless networks and ad hoc wireless networks.

Network games have been studied in the context of road traffic since
the 1950s, when Wardrop proposed the following definition for a
stable traffic flow on a transportation network: ``The journey times
on all the routes actually used are equal, and less than those which
would be experienced by a single vehicle on any unused route'' (see
p. 345 of \cite{wardrop-ice-1952}). In fact, this definition of
equilibrium is different from the one proposed by Nash. Expressing
the NE in terms of network flows, one can say that a network flow
pattern is an NE if no individual decision maker in the network can
change to a less costly strategy or route. When the number of
decision makers in the game is finite, an NE can be achieved without
the costs of all used routes being equal, contrary to Wardrop
equilibrium (WE) principle. The WE assumes therefore that the
contribution to costs or delays by any individual user is zero. In
other words, the population of users is considered infinite. In some
cases, Wardrop principle represents a limiting case of the NE
principle as the number of users becomes very large
\cite{haurie-net-1985}. Since routing in networks is also a
fundamental part of communication networks, it was natural to expect
the concepts of WE and of NE to appear in routing games. The natural
analogy between packets in a network and cars on the road was not
satisfactory for applying the Wardrop concept to networking, since
packets, unlike cars, do not choose their routes. These are
determined by the network routing protocols. Moreover, for a long
time, routing issues did not concern wireless networking as it was
restricted to the access part to a network where each terminal is
associated with a given BS.

\begin{figure}
  \begin{center}
  \includegraphics[angle=90,width=17cm]{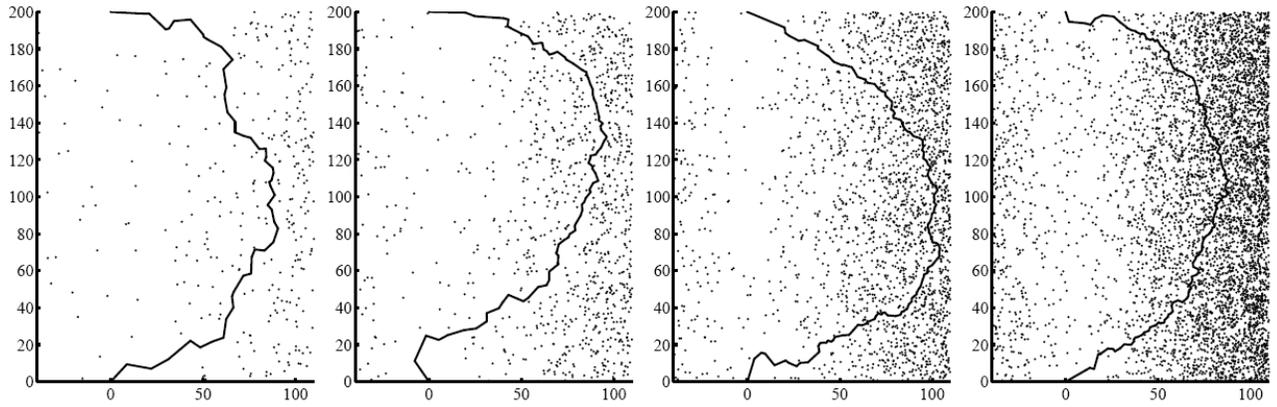}
    \caption{Minimum cost routes in increasingly large networks. Optimal routes tend to
    follow to the Fermat principle for light propagation in non-homogeneous media.}
    \label{fig:dense-net}
\end{center}
\end{figure}

Routing has become relevant to wireless networking when ad-hoc
networks have been introduced, see e.g., the pioneering work by
Gupta and Kumar \cite{gupta-cdc-1997}. Yet game theory has
penetrated into ad-hoc networks in a completely unexpected way.
Indeed, to our knowledge, it is in this framework that
non-cooperative game theory appears for the first time as a basic
powerful tool to solve problems that do not involve competition. In
\cite{gupta-cdc-1997}, Gupta and Kumar propose a routing algorithm
for packets that should satisfy some properties: packets should be
routed by the network so as to follow shortest paths. It is thus not
the average delay in the network that is minimized; the design
objectives are to find routing strategies that result in (i)
equalizing delays of packets that have the same source and
destination so as to avoid re-sequencing delays that are quite
harmful for real time traffic as well as for the TCP protocol for
data  transfers, and (ii) making these delays minimal, i.e., not to
use routes that have delays larger than the minimal ones. The
authors design a network routing protocol that achieves these two
objectives. But these objectives are nothing else than the
WE definition, a definition that arises in a completely
non-competitive context. It is only ten years later that the WE
penetrates into ad-hoc networks in a competitive game context. The
problem of competitive routing in massively dense ad-hoc networks
has been introduced in \cite{jacquet-mobihoc-2004}. The author
observes that as the density of mobiles grow, shortest paths tends
to converge to some curves that can be described using tools from
geometrical optics in non-homogeneous media (see Fig.
\ref{fig:dense-net}). Later, the electromagnetic field has been used
to approximate the routes of packets in massively dense ad-hoc
networks (see \cite{toumpis-interperf-2006} and references therein).
Only later, in \cite{silva-interperf-2008}\cite{altman-icanw-2008}
the natural concept of WE was introduced in massively dense ad-hoc
networks as a natural tool that deals with the competitive
interactions between mobiles. This is a special version of WE
defined on a continuum limit of a network, whose links and nodes are
so dense that they are replaced by a plain.

\subsection{Beyond static games}

We have already mentioned  in Sec. \ref{sec:selection} the important
fact that Nash equilibria predicted in a static game with complete
information can be the effectively observed outcome of a game (with
partial information) played in several steps where each player
observes the actions played by the others, react to them by playing
his BR, the others update their strategies accordingly, and so on.
This clearly establishes a first link between static games and
multi-step games. It turns out that the static game analysis is also
very important to study a certain class of dynamic games namely
repeated games, which consists, in their standard formulation
\cite{aumann-report-1966}, in repeating the same game at every time
instance. The players want to maximize their utility averaged over
the whole game duration. Interestingly, equilibria in certain
repeated games can be predicted from the analysis of the game played
in one shot. An important result in infinitely repeated games is the
folk theorem, which provides the set of possible Nash equilibria of
the repeated game (see e.g., \cite{aumann-book-1981} for more details).

Another important class of games are evolutionary games
\cite{maynardsmith-nature-1973}. A concept of solution for these
games is the concept of equilibria evolutionary stable (EES). This
concept generalizes the NE in the sense that these equilibria are
stable to the deviation of a fraction of the whole population (this
therefore represents a large number). Additionally, these types of
games are less demanding in terms of rationality assumptions.

\section{Conclusion}

Even in this short article, we have seen that game theory offers
many strong results to prove existence of equilibria, select one of
them when there are multiple equilibria and improve their
efficiency. Note, however, that the current literature comprises not
so many results to tackle the equilibrium uniqueness issue. In any
case, all the provided results are based on certain assumptions in
terms of information, rationality, and game stationarity. A real
challenge for wireless games will be to be able to characterize
equilibria under more realistic assumptions. For example, it is
well known that CSI is always imperfect and that rationality can be
an arguable assumption in heterogeneous wireless networks. And what
about stationarity in wireless networks where channels can vary
rapidly and the number of players can change? A real equilibrium
theory specific to wireless networks will probably need to be built
to address these issues. From the practical point of view, the learning
approach in wireless networks opens a large avenue for technological
innovation and designing algorithms implementable in real life wireless networks.

\bibliography{biblio}

%%%%%%%%%%%%%%%%%%%%%%%%%%%%%%%%%%%%%%%%%%%%%%%%%%%%%%%%%%%%%%%%%%%%%%%%%%%%%
%%%%%%%%%%%%%%%%%%%%%%%%%%%%%%%%%%%%%%%%%%%%%%%%%%%%%%%%%%%%%%%%%%%%%%%%%%%%%
%%%%%%%%%%%%%%%%%%%%%%%%%%%%%%%%%%%%%%%%%%%%%%%%%%%%%%%%%%%%%%%%%%%%%%%%%%%%%
%%%%%%%%%%%%%%%%%%%%%%%%%%%%%%%%%%%%%%%%%%%%%%%%%%%%%%%%%%%%%%%%%%%%%%%%%%%%%

\end{document}